\documentclass[amsmath,amssymb,twocolumn,letter]{jpsj3} 
\usepackage{bm}
\usepackage{graphicx,color}%

\title{Stability of Unconventional Superconductivity on Surfaces of Topological Insulators}
\author{  
Yuto \textsc{Ito}$^{1}$, Youhei \textsc{Yamaji}$^{1,2}$\thanks{present address: Department of Physics and Astronomy, Rutgers University, 136 Frelinghuysen Road, Piscataway, NJ 08854-8019, USA}, and Masatoshi \textsc{Imada}$^{1,2}$}

\inst{
$^{1}$ Department of Applied Physics,  
 University of Tokyo, 7-3-1 Hongo, Bunkyo-ku, Tokyo 113-8656 \\  
$^{2}$ Japan Science and Technology Agency, CREST, Honcho, Kawaguchi, Saitama 332-0012, Japan 
}
\date{today}

\abst
{
{
Superconductivity on the surface of topological insulators is known to be anisotropic and unconventional in that the symmetry is the mixture of $s$-wave and nodeless $p$-wave component.}
In contrast to Anderson's theorem for the insensitivity of the $s$-wave superconducting critical temperature to the nonmagnetic (time-reversal symmetric (TRS)) impurities, anisotropic superconductors {including nodeless $p$-wave one} are in general fragile even with small concentration of the TRS impurities. 
{
By employing the {Abrikosov}-Gor'kov theory, we clarify that this type of unconventional superconductivity emergent on the surface state of the strong topological insulators robustly survive against TRS impurities.
}
}
\kword{topological insulator, helical Dirac electron, unconventional superconductivity, impurity scattering, time reversal symmetry}
\begin{document}
\maketitle
 {Topological insulator(TI) is} a new quantum state of matter.\cite{cit:PRL95_146802,cit:PRL98_106803,cit:PRB75_121306,cit:PRB79_195322}
TIs are fully gapped in bulk as ordinary insulators but also have topologically protected conducting states on their boundaries. 
For example, two-dimensional (2D) TIs have one-dimensional ballistic conducting states and this conducting states are found in HgTe quantum wells \cite{cit:Science318_766}.
Moreover, a class of three-dimensional (3D) TI namely, strong TI,\cite{cit:PRL98_106803}
 has been predicted to
have metallic 2D surface states, which
have recently been observed by angle-resolved photoemission spectroscopy\cite{cit:Nature452_970,cit:Nphys5_398}.
Such surface states consist of so-called helical Dirac electrons, which occupy a single
spin state (pointing specific direction $\vec{s}\ $) depending on the electrons' momentum $\vec{p}$ as
$\vec{s}\propto\vec{p}$.

One of the most notable properties of strong TI is the robustness of the metallic surface states to TRS impurities.
This robustness originates from the absence of back-scattering processes when the system has the time reversal symmetry.
In the non-interacting electron systems, back-scattering processes due to impurities play important roles in forming localized states that do not contribute to the electric current at zero temperature\cite{cit:PR109_1492}.
In 2D electron systems, such localization leads to insulating states at zero temperature in the thermodynamic limit\cite{cit:PRL42_1979}.
However, when back-scattering processes are prohibited from some symmetric reasons, metallic states persist.

In other words, weak anti-localization is observed in TIs when we focus on weakly localized electron systems.
Impurity scatterings usually induce
negative quantum corrections to conductivities.
However, in systems with a spin-orbit coupling (SOC) such as
strong TIs, the quantum correction to the conductivity 
becomes positive and
results in weak anti-localization effects\cite{cit:PTP63_707}.
Furthermore when Dirac electrons, which are not necessarily helical, are scattered by impurities, accumulated $\pi$ Berry phase around Dirac cone yields weak anti-localization\cite{cit:JPSJ67_2857}.
In the most famous Dirac electron system, graphene with four Dirac cones, there are inter-valley scatterings, which lead to localization in the presence of disorders.
In contrast to the graphene,
helical Dirac electrons
on surfaces of strong TIs
are composed of odd numbers of Dirac cones and they show robust anti-localization\cite{cit:PRL98_106803}.

{
Recently possible anisotropic unconventional superconductivity (SC) with the mixture of $s$-wave and nodeless $p$-wave component in TIs has been studied.
It is known that even an $s$-wave attractive interaction necessarily induces the pairing potential composed of the mixture of $s$-wave and {$p$-wave} component on the surface of TI.\cite{cit:PRB81_184502}
Moreover by proximity effect with $s$-wave superconductors, SC with the same symmetry above is induced on the surface of TI in contact with the conventional superconductors when the Fermi energy is away from Dirac point}.\cite{cit:PRL100_096407,cit:PRB81_241310}

{Such unconventional SC attracts much attention from viewpoints of applications.}
The unconventional SC on the surface of TI is proposed to apply for quantum computations using Majorana fermions caused by the proximity effect between a superconductor and the surface states of TI.\cite{cit:PRL100_096407}
Introducing SC on the surfaces of TI has been a challenge for experimental researches\cite{cit:PRL104_057001}. 

A fundamental property of SC is their impurity effects. It is well known that the $s$-wave SC is robust to the TRS impurities\cite{cit:JPhysChemSolids11_26} while anisotropic unconventional SCs such as $d$-wave or $p$-wave SC are fragile to TRS impurities in general.\cite{cit:PRB37_4975,cit:PRB48_653,cit:PR131_1553,cit:RMP} A simple extension also leads to the fragility of the {nodeless $p$-wave} as well. 

{
Although the unconventional SC on the surface of TI have $s$-wave pairing potential component, they necessarily have $p$-wave component with the same amplitude. 
Therefore the response to TRS impurity concentration can be different from ordinary $s$-wave SC.
Moreover due to the anti-localization which helical Dirac electrons show, the impurity effect on this type of SC is nontrivial.
}
In this letter, in contrast to the widely accepted fragilities of the unconventional superconductors, we reveal that an unconventional SC stabilized on the surface of TI is robust to the TRS impurity scattering. This conclusion is obtained by analyzing the dependence of the critical temperature $T_{\rm c}$ on the TRS impurity concentration by the Abrikosov-Gor'kov (AG) theory{\cite{cit:JETP12_1243}}.

In order to study how TRS impurities affect the SCs on the surface state of strong TI, we introduce an effective model of the surface state.
Our model is a 2D helical Dirac electron system with an $s$-wave attractive interaction and in the presence of small concentration of impurities that cause on-site TRS scatterings. 
Therefore our Hamiltonian consists of three parts i.e. a 2D helical Dirac electron dispersion, $H_0$, an $s$-wave attractive interaction term, $H_{\rm int}$ and an on-site TRS impurity scattering term, $H_{\rm imp}$:
\begin{eqnarray}
H = H_0 + H_{\rm int} + H_{\rm imp}.
\end{eqnarray}

The 2D helical Dirac electron  dispersion {$H_0$}
is written as 
\begin{eqnarray}
H_0 = \sum_{\bm k} c^\dagger(\bm k)(v_F \bm\sigma\cdot\bm k-\mu I)c(\bm k)
\end{eqnarray}
with {the} Fermi {velocity} $v_F>0$.
In this paper, we set the Fermi energy $\mu>0$ for simplicity.
At $\mu=0$ case, we have to consider two branches of Dirac electrons because the both branches contribute to SC.
Moreover, when $\mu$ is closer to $0$, the SC is harder to be observed because the density of states at the Fermi energy gets fewer.
Pauli matrix $\bm\sigma$ describes electron's spin and $c(\bm k)=(c_{\bm k\uparrow}\ c_{\bm k\downarrow})^T$.
We consider the 2D Dirac Hamiltonian of the conduction electrons with only one Dirac cone
for simplicity.
We note that a model with single Dirac cone is sufficient to understand the essential
properties of the surface states of strong TIs.
By using a unitary transformation 
$d^\dagger_{\bm k,\tau} = (c^\dagger_{\bm k\uparrow} + \tau e^{i\theta_{\bm k}}c^\dagger_{\bm k \downarrow})/{\sqrt{2}}\ (\tau = +\ {{\rm or}}\ - )$
, this term is diagonalized as
\begin{eqnarray}
H_0 = \sum_{\bm k} d^\dagger({\bm k}) (v_F|\bm k|\tau_z-\mu)d ({\bm k}),     \label{Eq3}
\end{eqnarray}
where $\tau_z$ is Pauli $z$ matrix describing branches of Dirac electrons and $d(\bm k)=(d_{\bm k+}\ d_{\bm k-})^T$.
The index $\tau$ ($\tau = \pm$) represents
branches of Dirac electrons{. Here the} ``+" (``-") branch represents the branch above (below) Dirac point.
In the unitary transformation, an angle parameter $\theta_{\bm k} = \arg(k_x+ik_y)$ is introduced.
The operators $d$ and $d^\dagger$ satisfy the
anti-commutation relation $\{d_{\bm k\tau},d_{\bm k'\tau'}^\dagger \} = \delta_{\bm k\bm k'}\delta_{\tau\tau'}$.

Then we introduce an $s$-wave attractive interaction $H_{\rm int}$ with an energy cutoff.
For example, the attractive interaction may originate from an electron-phonon coupling. 
We assume that $H_{\rm {int}}$ is written as
\begin{eqnarray}
H_{\rm int}
= 
\frac{1}{2}\sum_{\bm k,\bm k',s,s'} V_{\bm k\bm k'ss'}
c_{-\bm ks}^\dagger c_{\bm ks'}^\dagger 
c_{\bm k's'}c_{-\bm k's}.
\end{eqnarray}
In this equation, we assume that
\begin{eqnarray}
V_{\bm k\bm k'\downarrow\uparrow} =V_{\bm k\bm k'\uparrow\downarrow} =
\left\{
\begin{array}{cc}
-g&
(
\xi_{\bm k},\xi_{\bm k'}\in[-\omega_c,\omega_c])\\
0&
(
\xi_{\bm k},\xi_{\bm k'}\not\in[-\omega_c,\omega_c])
\end{array}
\right.,
\end{eqnarray}
\begin{eqnarray}
V_{\bm k\bm k'\uparrow\uparrow} =V_{\bm k\bm k'\downarrow\downarrow} =
\left\{
\begin{array}{cc}
-g'&
(
\xi_{\bm k},\xi_{\bm k'}\in[-\omega_c,\omega_c])\\
0&
(
\xi_{\bm k},\xi_{\bm k'}\not\in[-\omega_c,\omega_c])
\end{array}
\right.,
\end{eqnarray}
with the cutoff $\omega_c  \ll \mu$, and $g, g' > 0$.
Here
we define the energy $\xi_{\bm k}$ as the energy of the ``+" branch measured
from the Fermi energy as $\xi_{\bm k} = v_F|\bm k|-\mu$.
We note that, from the condition $\omega_c \ll \mu$, the interaction only affects electrons on the ``+" branch.
Therefore, we neglect ``-" branch and we write
$d_{+}^{(\dagger)}$ as $d^{(\dagger)}$ below.
Then the interaction term is written as
\begin{eqnarray}
H_{int} \simeq
-\frac{g}{4}\sum_{\bm k,\bm k'} ^*
e^{i(\theta_{\bm k'}-\theta_{\bm k})}d^\dagger_{-\bm k}d^\dagger_{\bm k}d_{\bm k'}d_{-\bm k'} .
\end{eqnarray}
The summation $\displaystyle\sum^*_{\bm k,\bm k'}$ represents that taken in the region $\xi_{\bm k},\xi_{\bm k'}\in[-\omega_c,\omega_c]$.
The terms including $V_{\bm k\bm k'\uparrow\uparrow}$ and $V_{\bm k\bm k'\downarrow\downarrow}$ vanish because they must be odd functions of $\bm k$.
On the other hand, the terms including $V_{\bm k\bm k'\uparrow\downarrow}$ and $V_{\bm k\bm k'\uparrow\downarrow}$ survive because they are even functions of $\bm k$.

Here we introduce an on-site TRS impurity scattering term as
\begin{eqnarray}
H_{\rm imp}
=
\frac{u}{S}\sum_{i=1}^{N_{\rm i}}\sum_{\bm k,\bm q}e^{-i\bm q\cdot\bm R_i}
c^\dagger(\bm k+\bm q)c({\bm k}),
\end{eqnarray}
where
$N_{\rm i}$ is the number of the impurities in the system, $S$ is the size of the system, and $\bm R_i\ (i = 1,\cdots,N_{\rm i})$ is the impurity location.
Because on-site impurity scatterings are assumed here, 
scattering amplitudes do not depend on momentum transfer $\bm q$.
By using the unitary transformation introduced above eq.(\ref{Eq3}), eq.(8) leads to 
\begin{eqnarray}
H_{\rm imp}
=
\frac{u}{S}
\sum_{i=1}^{N_{\rm i}}
\sum_{\bm k,\bm q}e^{-i\bm q\cdot\bm R_i}
P(\theta_{\bm k}-\theta_{\bm k+\bm q})
d^\dagger_{\bm k+\bm q} d_{\bm k},
\end{eqnarray}
where $P(\theta)=e^{i\theta/2}\cos(\theta/2)$ is a phase factor {specific} to Dirac electron systems.
The phase factor $P(\theta_{\bm k}-\theta_{\bm k+\bm q})$ contributes to
the $\pi$ Berry phase and leads to anti-localization effect of single Dirac cone systems. \cite{cit:JPSJ67_2857}

We introduce a SC paring potential and write down the BCS mean-field Hamiltonian.
By introducing a pair potential
\begin{eqnarray}
\Delta(\bm k)   = e^{-i\theta_{\bm k}} g\sum_{\bm k'}' e^{i\theta_{\bm k'}} \langle d_{\bm k'} d_{-\bm k'} \rangle,\label{Eq10}
\end{eqnarray}
our interaction term is approximated as 
\begin{eqnarray}
\mathcal{H}_{\rm int} \simeq -\sum_{\bm k}' \left[\Delta(\bm k)d^\dagger_{-\bm k}d^\dagger_{\bm k} + \Delta^*(\bm k)d_{\bm k}d_{-\bm k} \right], 
\end{eqnarray}
where the summation $\displaystyle{\sum'_{\bm k}}$ represents that taken in the momentum space such that $\xi_{\bm k}\in[-\omega_c,\omega_c]$ and $k_x>0$.
We linearize the gap equation in terms of $\Delta$ because we only consider $T_{\rm c}$.
The pair potential represented by eq.(\ref{Eq10}) is valid when we can neglect the mixture of ``+" and ``-" branches.
This pair potential depends on the momentum $\bm k$ and its dependence is represented as 
\begin{eqnarray}
\Delta(\bm k) = \Delta e^{-i\theta_{\bm k}}.
\end{eqnarray}

Therefore the equation $\Delta(\bm k) = -\Delta(-\bm k)$ holds.
This pair potential resembles one appearing in the spinless chiral $p$-wave superconductor.\cite{cit:PRB61_10267}

This SC order is unconventional because it is composed of a mixture of two different pairing symmetries.
The pairing potential matrix in terms of the original electrons operator $c$ is defined from the equation 
\begin{eqnarray}
\mathcal{H}_{\rm int} = -\sum_{\bm k}' \sum_{s_1,s_2} \left[c_{-\bm ks_1}^\dagger \hat\Delta_{s_1s_2}(\bm k) c_{\bm ks_2}^\dagger + h.c. \right],
\end{eqnarray}
where
\begin{eqnarray}
\hat\Delta(\bm k)
\nonumber&=&
\hat\Delta_{s}(\bm k)+\hat\Delta_t(\bm k)\\
\hat\Delta_{s}(\bm k) &=& \frac{\Delta}{2}
\begin{pmatrix}
0 & 1 \\ -1 & 0
\end{pmatrix}
,
\hat\Delta_t(\bm k)=\frac{\Delta}{2}
\begin{pmatrix}
e^{-i\theta_{\bm k}} & 0 \\ 0 & -e^{i\theta_{\bm k}}
\end{pmatrix}.
\end{eqnarray}
Therefore the SC induced by $H_{\rm int}$ is the mixture of singlet $s$-wave pairing and triplet
{nodeless $p$-wave} pairing with the same amplitude. \cite{cit:PRB81_184502}
The $s$-wave pairing term $\hat \Delta_s$ has no $\bm k$ dependence while the {$p$-wave} term $\hat \Delta_t$ has a phase determined from the  direction of $\bm k$.

In order to analyze $T_{\rm c}$, we construct Gor'kov equations and linearize the gap equation to determine $T_{\rm c}$.\cite{cit:JETP12_1243} 
Because the long-range SC order does not develop in 2D systems at finite temperature, $T_{\rm c}$ calculated from the mean-field theory provides a criteria of the Berezinskii-Kosterlitz-Thouless (BKT) transition for development of the quasi-long-range order\cite{cit:JETP34_610,cit:JPhysC 6_1181}.

First, we introduce two thermodynamic Green's functions 
$
G_{\bm k,\bm k'}(\tau) = -\langle T_\tau d_{\bm k}(\tau)d^\dagger_{\bm k'}\rangle\ \mathrm{and}\ 
F^*_{\bm k,\bm k'}(\tau)=\langle T_\tau d^\dagger_{-\bm k}(\tau)d_{\bm k'}^\dagger\rangle,
$
where $G$ ($F^*$) is the normal (anomalous) Green's function, which describes the dynamics of Cooper pairs.  
From the equations of motion for two Green's functions, we can construct the Gor'kov equations as
\begin{eqnarray}
&&\delta_{\bm k,\bm k'} = (i\omega_n-\xi_{\bm k})G_{\bm k,\bm k'}(i\omega_n) + \Delta(\bm k)F^*_{\bm k,\bm k'}(i\omega_n)\nonumber\\
&-&\frac{u}{S}\sum_{i=1}^{N_{\rm i}}\sum_{\bm q}G_{\bm q,\bm k'}(i\omega_n)e^{i(\bm q-\bm k)\cdot\bm R_i}
P(\theta_{\bm q}-\theta_{\bm k}),  \label{Eq16}
\end{eqnarray}
and
\begin{eqnarray}
&&0 = (i\omega_n+\xi_{\bm k})F^*_{\bm k,\bm k'}(i\omega_n) + \Delta^*(\bm k)G_{\bm k,\bm k'}(i\omega_n)\nonumber\\
&+&\frac{u}{S}\sum_{i=1}^{N_{\rm i}}\sum_{\bm q}F^*_{\bm q,\bm k'}(i\omega_n)e^{i(\bm q-\bm k)\cdot\bm R_i}
P(\theta_{\bm k}-\theta_{\bm q}),
\end{eqnarray}
where $\omega_n = (2n+1)\pi T$ is the fermionic Matsubara frequency, and $T$ is the temperature. 
Moreover, we assume that
$\Delta(\bm k)$ has a value independent of the frequency when $\xi_{\bm k}\in[-\omega_c,\omega_c]$.
By linearizing in terms of $\Delta$, eq.(\ref{Eq16}) leads to
\begin{eqnarray}
&&\delta_{\bm k,\bm k'} = (i\omega_n-\xi_{\bm k})G_{\bm k,\bm k'}(i\omega_n)\nonumber \\
&-&\frac{u}{S}\sum_{i=1}^{N_{\rm i}}\sum_{\bm q}G_{\bm q,\bm k'}(i\omega_n)e^{i(\bm q-\bm k)\cdot\bm R_i}
P(\theta_{\bm q}-\theta_{\bm k}). \label{Eq9}
\end{eqnarray}
By using a perturbation series expansion with respect to
$u/S$, the Green's function $G$ is represented as
\begin{eqnarray}
G_{\bm k,\bm k'}(i\omega_n) = \sum_{n=0}^{\infty} G^{(n)}_{\bm k,\bm k'}(i\omega_n)
\end{eqnarray}
where
$$G^{(0)}_{\bm k,\bm k'}(i\omega_n) = \delta_{\bm k,\bm k'}G^{0}_{\bm k}(i\omega_n)$$
and
\begin{eqnarray}
&&G^{(n)}_{\bm k,\bm k'}(i\omega_n) 
\nonumber= G^{0}_{\bm k}(i\omega_n)
\frac{u}{S}\sum_{\bm q}\sum_{i=1}^{N_{\rm i}}e^{i(\bm q-\bm k)\cdot\bm R_i}\\
&\times&
P(\theta_{\bm q}-\theta_{\bm k})G^{(n-1)}_{\bm q,\bm k'}(i\omega_n)\ (n=1,2,\cdots).
\end{eqnarray}
The non-perturbative Green's function $G^{0}_{\bm k}(i\omega_n) = (i\omega_n-\xi_{\bm k})^{-1}$ is introduced above.
By substituting eq.(18) and eq.(19) into eq.(16) and using eq.(17), the anomalous Green's function is obtained as
{\begin{eqnarray}
F^*_{\bm k,\bm k'}(i\omega_n) &=& \sum_{\bm q}\Delta^*(\bm q)
G_{-\bm q,-\bm k}(-i\omega_n)G_{\bm q,\bm k'}(i\omega_n). \label{Eq12}
\end{eqnarray}}
On the other hand, from eq.(\ref{Eq10}), we obtain 
\begin{eqnarray}
\Delta^*
= gT\sum_{i\omega_n}\sum'_{\bm k'}e^{-i\theta_{\bm k'}}F^*_{\bm k',\bm k'}(i\omega_n), \label{Eq13}
\end{eqnarray}
where  $\Delta^*(\bm k)=\Delta^*e^{i\theta_{\bm k}}$.
From eqs.(\ref{Eq12}) and (\ref{Eq13}), we obtain the equation to determine $T_{\rm c}$
as follows:
\begin{eqnarray}
(gT_c)^{-1} \nonumber
&=&
\sum_{n,m}
F(n,m)\\
F(n,m) &=& \nonumber
\sum_{i\omega_n}\sum_{\bm k}\sum_{\bm k'}'
e^{i(\theta_{\bm k}-\theta_{\bm k'})}\\
&\times&
G^{(n)}_{-\bm k,-\bm k'}(-i\omega_n)G^{(m)}_{\bm k,\bm k'}(i\omega_n)%
.\label{Eq22}
\end{eqnarray}

Then we analyze dependence of $T_{\rm c}$ on {the} impurity concentration $n_i$
$=N_{\rm i}/S$ by
using the perturbative AG theory,
where $n_i$ is the small parameter\cite{cit:JETP12_1243}.
According to the AG theory, we neglect spatial correlations between different impurities.
This approximation i.e., the impurity average, is justified when the impurity concentration is small.
By the impurity average operation, the term
$ \displaystyle{
\frac{1}{S}\sum_{i,j=1}^{N_{\rm i}} e^{i\bm q\cdot(\bm R_i-\bm R_j)}}
$
is replaced by
$
n_i,
$
and only $\mathcal{O}(n_i)$ terms are retained.
We perform
the impurity average of the right hand side of eq.(\ref{Eq22}) for each set of $(n,m)$, step by step, as follows:
We only retain the terms in eq.(\ref{Eq22}) that satisfy $n,m\le2$ and $n+m\le2$ because
these terms are of the lowest order in the $n_i$-dependence of $T_{\rm c}$.
The non-perturbative term i.e., $F(0,0)$, which is illustrated in Fig. 1(a),
is calculated as,
\begin{eqnarray}
F(0,0)  
=\sum_{i\omega_n}\sum_{\bm k}' G^{0}_{\bm k}(i\omega_n)G^{0}_{-\bm k}(-i\omega_n).
\end{eqnarray}
The phase factor vanishes because only the term satisfying $\bm k = \bm k'$ remains.
The term $F(0,1)\ \mathrm{and}\ F(1,0)$
shown in Fig. 1(b) are negligible for the estimation of $T_{\rm c}$
because these terms just cause a constant self-energy shift and do not contribute to
relaxation processes due to impurity scatterings.
Then $F(1,1)$, which corresponds to the diagram
illustrated in Fig. 1(c), is represented as
\begin{eqnarray}
&&F(1,1) = \frac{n_iu^2}{S}\sum_{i\omega_n}\sum_{\bm k,\bm k'}' G^{0}_{\bm k}(i\omega_n)G^{0}_{\bm k'}(i\omega_n)\nonumber \\
&\times&G^{0}_{-\bm k}(-i\omega_n)G^{0}_{-\bm k'}(-i\omega_n).
\end{eqnarray}
The diagram illustrated in Fig. 1(d) represents $F(0,2)$ or $F(2,0)$, and is given by
\begin{eqnarray}
&&F(0,2)\nonumber =F(2,0)\\ \nonumber&=& 
\frac{n_iu^2}{S}\sum_{i\omega_n}\sum_{\bm k}'\sum_{\bm q}\cos^2\left(\frac{\theta_{\bm k}-\theta_{\bm q}}{2}\right)
G^0_{\bm q}(i\omega_n)\\
&\times&G^0_{\bm k}(i\omega_n)G^0_{\bm k}(i\omega_n)G^0_{-\bm k}(-i\omega_n).
\end{eqnarray}
\begin{figure}[ht]
\begin{center}
\includegraphics[width=5cm]{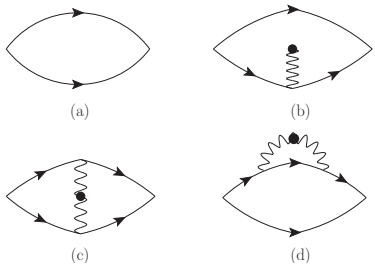}
\end{center}
\caption{Diagrams for terms of the lowest order in eq.(\ref{Eq22}). (a), (b), (c) and (d) correspond to $(n,m)=(0,0)$, $(n,m)=(0,1)\ \mathrm{or} (1,0)$,
$(n,m)=(1,1)$ and $(n,m)=(0,2)\ \mathrm{or}\ (2,0)$, respectively.
Wavy lines represent impurity scatterings.}
\end{figure}
To combine these terms, $F(0,0), F(1,1), F(2,0)$, and $F(0,2)$, we obtain 
\begin{eqnarray}
1=gT_c \sum_{i\omega_n}\sum_{\bm k}' 
\frac{1+(2\tau_2|\omega_n|)^{-1}}{\omega_n^2\left[1+(2\tau_1|\omega_n|)^{-1}\right]^2 + \xi_{\bm k}^2},
\end{eqnarray}
where $\tau_1$ and $\tau_2$ are relaxation times which are related to $F(2,0)$ and $F(1,1)$, respectively.
Both quantities are calculated as
\begin{eqnarray}
\frac{1}{2\tau_1} = \frac{\pi n_i u^2}{S}\sum_{\bm k}\cos^2(\theta_{k}/2)\delta(\xi_{k})=\frac{1}{2}\pi n_i u^2 N_0,\label{tau_1_inv}
\end{eqnarray}
and
\begin{eqnarray}
\frac{1}{2\tau_2} = \frac{\pi n_i u^2}{S} \sum_{\bm k}' \delta(\xi_q)=\frac{1}{2}\pi n_i u^2 N_0 ,
\label{tau_2}
\end{eqnarray}
where
$N_0 = \sum_{k}\delta(\xi_k)/S$ is the density of states at Fermi energy.
According to the AG theory, when the concentration $n_i$ is small, $T_{\rm c}$  is obtained from the equation.\cite{cit:JETP12_1243}
\begin{eqnarray}
T_c(n_i) = T_{c}(0) - \frac{\pi}{4\tau_s(n_i)},
\end{eqnarray}
where
$\Psi$ is the digamma function.
In these equations, $T_{\rm c}$ without impurities is
$\displaystyle T_c(0) = \frac{2e^{\omega_c}}{\pi }e^{-\frac{2}{gN_0}}$
and $\tau_s$ is obtained from
$(\tau_s)^{-1} = (2\tau_1)^{-1} - (2\tau_{2})^{-1}.$
Then, from eqs.(\ref{tau_1_inv}) and (\ref{tau_2}), we obtain
$
(\tau_s)^{-1} = 0, 
$
up to the lowest order of $n_i$, and
we conclude at least for a small impurity concentration,
\begin{eqnarray}
T_c(0)- T_c(n_i) = \mathcal{O}(n_i^{2}).
\end{eqnarray}
Such cancellation of the relaxation times does not occur in the case of $d$-wave\cite{cit:PRB37_4975,cit:PRB48_653} or $p$-wave\cite{cit:PR131_1553,cit:RMP} SC orders.
The present SC order is anisotropic because it is composed of $s$-wave and {$p$-wave} order.
Therefore, our result supports that
anisotropic but robust SC orders can exist on the surface of TIs.

We have shown that unconventional SCs induced by the $s$-wave attractive interaction on surfaces of TIs
are robust to TRS disorders.
This conclusion is achieved by calculating dependence of $T_{\rm c}$  on TRS impurity concentration,
where $T_{\rm c}$ provides a criteria of the BKT transition for the quasi-long-range order in 2D.
Unconventional SC studied in the literature, such as the $d$-wave and chiral $p$-wave SC, is sensitively suppressed through scatterings by a tiny concentration of impurities because of the phase factor of the pairing potential.
In marked contrast, the unconventional SC on the surface of TI is robust because of the cancellation of two phase factors, one from the pairing potential and the other arising when a Dirac electron is scattered by a TRS impurity.
The robustness of the unconventional {nodeless $p$-wave} SC may favorably be tested in experiments.
{This result is applicable to SC on TI by proximity effect\cite{cit:PRL100_096407,cit:PRB81_241310} as well because it has the same Cooper pairing symmetry as the SC considered in this letter.}
An issue left for future is the case of anisotropic attractions, where SC order with different symmetries may occur.
\begin{acknowledgement}
The authors thank Moyuru Kurita, and Takahiro Misawa for fruitful discussions. Y. I. also thanks Masafumi Udagawa for valuable discussions. 
\end{acknowledgement}


\begin{thebibliography}{10}
	\bibitem{cit:PRL95_146802}C. L. Kane and E. J. Mele: Phys. Rev. Lett. \textbf{95} (2005) 146802.
	\bibitem{cit:PRL98_106803}L. Fu, C. L. Kane, and E. J. Mele: Phys. Rev. Lett. \textbf{98} (2007) 106803.
	\bibitem{cit:PRB75_121306}J. E. Moore and L. Balents: Phys. Rev. B \textbf{75} (2007) 121306.
	\bibitem{cit:PRB79_195322}R. Roy: Phys. Rev. B \textbf{79} (2009) 195322.
	\bibitem{cit:Science318_766}M. K\"{o}nig, S. Wiedmann, C. Br\"{u}ne, A. Roth, H. Buhmann, L. W. Molenkamp, X.-L. Qi, and S.-C. Zhang: Science \textbf{318} (2007) 766.
	\bibitem{cit:Nature452_970}D. Hsieh, D. Qian, L. Wray, Y. Xia, Y. S. Hor, R. J. Cava, and M. Z. Hasan: Nature \textbf{452} (2008) 970.
	\bibitem{cit:Nphys5_398}Y. Xia, D. Qian, D. Hsieh, L. Wray, A. Pal, H. Lin, A. Bansil, D. Grauer, Y. S. Hor, R. J. Cava, and M. Z. Hasan: Nat. Phys. \textbf{5} (2009) 398.
	\bibitem{cit:PR109_1492}P. W. Anderson: Phys. Rev. \textbf{109} (1958) 1492.
	\bibitem{cit:PRL42_1979}E. Abrahams, P. W. Anderson, D. C. Licciardello, and T. V. Ramakrishnan: Phys. Rev. Lett. \textbf{42} (1979) 673.
	\bibitem{cit:PTP63_707}S. Hikami, A. I. Larkin, and Y. Nagaoka: Prog. Theor. Phys. \textbf{63} (1980) 707.
	\bibitem{cit:JPSJ67_2857}T. Ando, T. Nakanishi, and R. Saito: JPSJ \textbf{67} (1998) 2857.
	{\bibitem{cit:PRB81_184502}L. Santos, T. Neupert, C. Chamon, and C. Mudry: Phys. Rev. B \textbf{81} (2010) 184502.}
	{\bibitem{cit:PRL100_096407}L. Fu and C. L. Kane: Phys. Rev. Lett. \textbf{100} (2008) 096407.}
	{\bibitem{cit:PRB81_241310}T. Stanescu, J. Sau, R. Lutchyn, and S. Das Sarma: Phys. Rev. B \textbf{81} (2010) 241310.}
	\bibitem{cit:PRL104_057001}Y. S. Hor, A. J. Williams, J. G. Checkelsky, P. Roushan, J. Seo, Q. Xu, H. W. Zandbergen, A. Yazdani, N. P. Ong, and R. J. Cava: Phys. Rev. Lett. \textbf{104} (2010) 057001.
	\bibitem{cit:JPhysChemSolids11_26}P. W. Anderson: J. Phys. Chem. Solids \textbf{11} (1959) 26.
	\bibitem{cit:PRB37_4975}A. J. Millis, S. Sachdev, and C. M. Varma: Phys. Rev. B \textbf{37} (1988) 4975.
	\bibitem{cit:PRB48_653}R. J. Radtke, K. Levin, H.-B. Sch\"uttler, and M. R. Norman: Phys. Rev. B \textbf{48} (1993) 653.
	\bibitem{cit:PR131_1553}R. Balian and N. R. Werthamer: Phys. Rev. \textbf{131} (1963) 1553.
	\bibitem{cit:RMP}M. Sigrist and K. Ueda: Rev. Mod. Phys. \textbf{63} (1991) 239.
	\bibitem{cit:PRB61_10267}N. Read and D. Green: Phys. Rev. B \textbf{61} (2000) 10267.
	\bibitem{cit:JETP12_1243}A. A. Abrikosov and L. P. Gor'kov: Sov. Phys. JETP \textbf{12} (1961) 1243.
	\bibitem{cit:JETP34_610}V. L. Berezinskii: Sov. Phys. JETP \textbf{34} (1972) 610.
	\bibitem{cit:JPhysC6_1181}J. M. Kosterliz and D. J. Thouless: J. Phys. C \textbf{6} (1973) 1181.
\end{thebibliography}
\end{document}